\documentclass[12pt,a4paper,final]{iopart}
\usepackage[usenames, dvipsnames]{color}

\usepackage{iopams}  
\usepackage{graphicx}
\usepackage[breaklinks=true,colorlinks=true,linkcolor=blue,urlcolor=blue,citecolor=blue]{hyperref}
\expandafter\let\csname equation*\endcsname\relax

\expandafter\let\csname endequation*\endcsname\relax
\usepackage{amsmath}

\begin{document}

\title{Active dynamics of tissue shear flow}

\author{Marko Popovi\' c$^{1}$, Amitabha Nandi$^{1,2}$, Matthias Merkel$^{1,3}$, Rapha\"el Etournay$^{4, 5}$, Suzanne Eaton$^{4}$, Frank J\"ulicher$^{1}$, Guillaume Salbreux$^{1,6}$}
\address{$^1$Max Planck Institute for the Physics of Complex Systems, N\"othnitzer Str. 38, 01187 Dresden, Germany}
\address{$^2$Department of Physics, Indian Institute of Technology Bombay, Powai, Mumbai 400076, India}
\address{$^3$Syracuse University, Syracuse, New York 13244, United States of America}
\address{$^4$Max Planck Institute of Molecular Cell Biology and Genetics, Pfotenhauerstr. 108, 01307 Dresden, Germany}
\address{$^5$Institut Pasteur, 25 rue du Dr. Roux, 75015 Paris, France}
\address{$^6$The Francis Crick Institute, 1 Midland Road, London NW1 1AT, United Kingdom}

\eads{\mailto{julicher@pks.mpg.de}, \mailto{guillaume.salbreux@crick.ac.uk}}

\begin{abstract}
We present a hydrodynamic theory to describe shear flows in developing epithelial tissues. We introduce hydrodynamic fields corresponding to state properties of constituent cells as well as a contribution to overall tissue shear flow due to rearrangements in cell network topology. We then construct a generic linear constitutive equation for the shear rate due to topological rearrangements and we investigate a novel rheological behaviour resulting from memory effects in the tissue. We identify two distinct active cellular processes: generation of active stress in the tissue, and actively driven topological rearrangements. We find that these two active processes can produce distinct cellular and tissue shape changes, depending on boundary conditions applied on the tissue. Our findings have consequences for the understanding of tissue morphogenesis during development.
\end{abstract}


\section{Introduction}\label{sec:introduction}
\begin{figure}
\centering
\includegraphics[width= .8\textwidth]{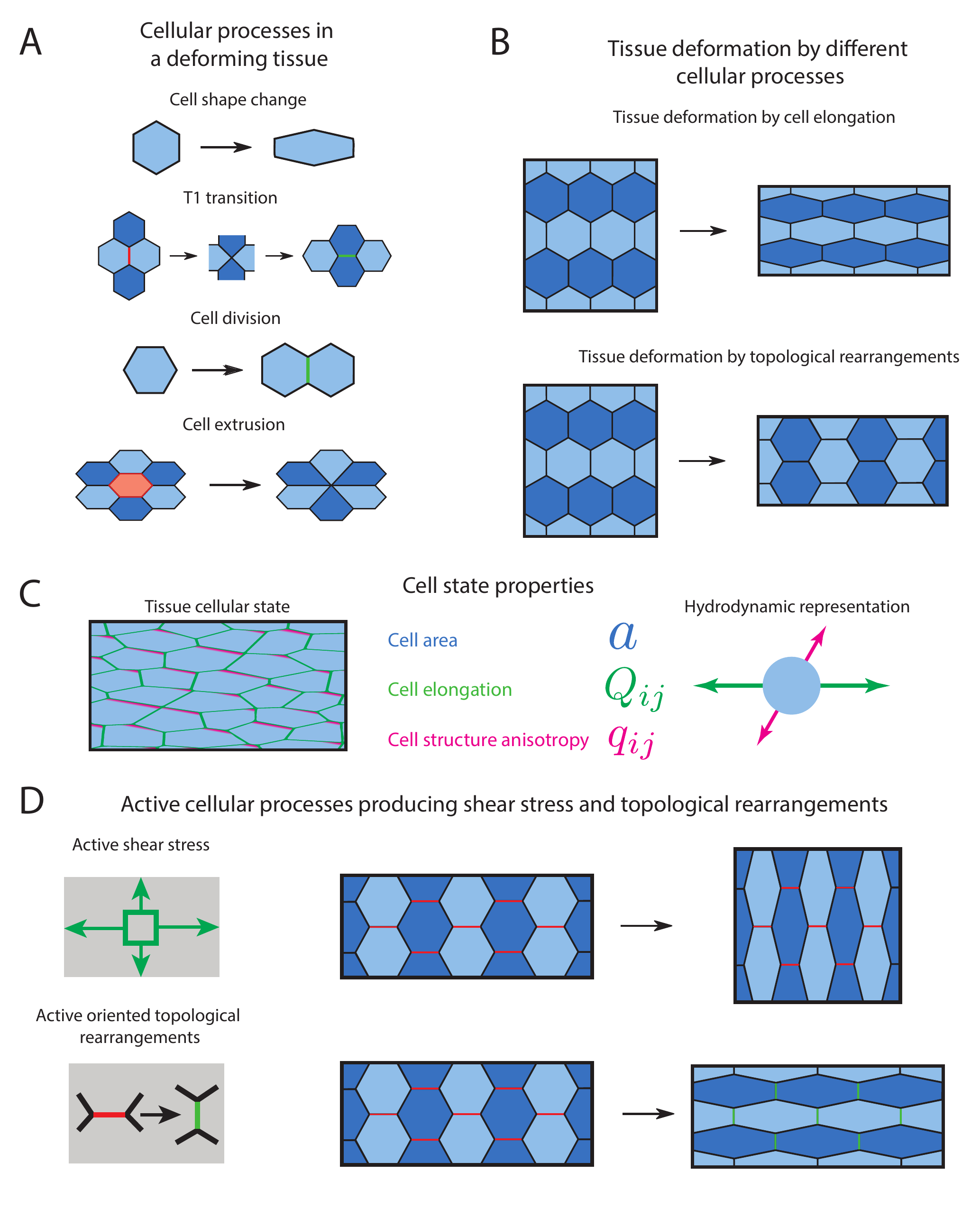} 
\caption{A) Cellular processes contributing to tissue deformation. Cell elongation: Deforming cells reshape tissue proportionally to cell area and elongation change. T1 Transition: During a T1 transition two 
cells lose a bond they share (red bond) and after passing through 4-fold vertex configuration (middle) a new bond (green) is created between the other two cells taking part in T1 transition. Cell division: Cell division produces two new daugher cells from a mother cells. Cell extrusion: During a cell extrusion a single cell (red) is removed from the tissue. 
B) Tissue deformation can arise from changes in cell elongation (top) and from topological rearrangements (bottom). Note that in the figure topological rearrangements were represented by T1 transitions but in general they can include both cell divisions and cell extrusions.
C) Cell state properties: cell area, cell elongation and structural anisotropy of cells are represented by hydrodynamic fields.
D) Active anisotropic processes in a tissue can result in anisotropic active stress (top),
or drive actively oriented topological rearrangements (bottom). Active stress and shear due to active oriented topological rearrangements produce different cell and tissue behavior (see Section \ref{sec:activityFlows}).}
\label{fig:cellularProcesses}
\end{figure}

During morphogenesis, epithelial tissues grow and reshape to form different organs in the adult animal. These tissue shape changes result from external stresses acting on the tissue as well as from autonomous force generation by cells \cite{LecuitLenne2007, Keller2012, HeisenbergBellaiche2013}. Cellular forces induce cell deformations and topological rearrangements of the network of bonds joining the cells. Topological rearrangements occurring in tissue morphogenesis include neighbour exchanges through $T_1$ transitions, cell divisions and cell extrusions. During $T_1$ transitions, an edge joining two cells shrinks and two neighbors loose their contacts, resulting in a 4-fold vertex (Figure \ref{fig:cellularProcesses} A). A new bond can then form, establishing a contact between two cells which were not neighbors before. During cell divisions, a new bond is formed between the two daughter cells, and during cell extrusion, an entire cell leaves the tissue (Figure \ref{fig:cellularProcesses} A). Topological rearrangements fluidify the epithelium through neighbour exchanges events, as has been observed in cell aggregates under compression \cite{marmottant2009role}, or through cell divisions and extrusion \cite{RanftFluidization}. In passive systems such as foams, topological rearrangements occur as a response to an external force deforming the system \cite{FoamFlowsReview, BianceLiquidFoams}. In biological tissues, which work out-of-equilibrium, topological rearrangements can however be internally driven by the system to generate deformation. In germ-band elongation of {\it Drosophila} embryos for instance, $T_1$ transitions are preferentially oriented, with cell bonds removed along the dorso-ventral axis of the tissue and added along the antero-posterior axis, and are actively driven by the cells \cite{LecuitNature, CollinetLennLecuit2015}, leading to tissue-scale reorganisation and flows. Similar actively driven oriented cell neighbor exchanges contribute to so called convergent extension processes that elongate the embryo along the body axis during development in many animals \cite{keller2000mechanisms, tada2012convergent}.

During tissue morphogenesis, tissue deformation can either arise from deformation of individual cells of the tissue, or from topological rearrangements of the tissue: for instance, cellular neighbour exchange can result in shear by rearranging the cell positions, without overall cell deformation  (Figure \ref{fig:cellularProcesses} B). The contribution of topological rearrangements to tissue flows has been measured to be a significant component of total tissue deformation in some morphogenetic processes in {\it Drosophila} \cite{tissueTectonics, wingMorpho, etournay2016tissueminer, Guirao2015}. It is however unclear how cell deformation, topological rearrangements and the overall tissue flow are physically coupled to each other. Similar decomposition approaches of the large scale material deformations were also used to study flows in foams \cite{marmottant2007elastic, graner2008discrete, marmottant2008discrete}.

The collective behaviour of cells in epithelia based on mechanics of single cells can be captured by different models. \cite{Honda, GranerPottsModel, Drasdo, farhadifar2007}. Vertex-model numerical simulations for instance attempt to capture the mechanics of epithelial tissues by representing the tissue by a network of bonds \cite{farhadifar2007}. The vertices are then subjected to forces derived from an effective mechanical energy, taking into account line tensions acting on the edges of the networks and an area elasticity. Cell-based simulations of tissues however require making specific hypotheses on cell mechanics. Alternatively, continuum theories have been proposed to describe tissue growth and deformation due to cell proliferation \cite{Bittig2008, RanftFluidization, Blanch-Mercader2014,  recho2016one, yeh2016hydrodynamics}, cell neighbor exchanges \cite{marmottant2009role} or both \cite{wingMorpho, tlili2015colloquium}. Collective migrations in cell monolayers have been studied using continuum models that couple the tissue mechanics with the tissue polarization and to the concentration of a chemical activator \cite{serra2012mechanical, kopf2013continuum, Banerjee2015stressWaves, notbohm2016cellular}. 

We have recently studied the cell deformations and tissue flows occurring during morphogenesis of the {\it Drosophila} pupal wing \cite{wingMorpho}. We found that topological rearrangements are oriented relative to the tissue axis and their preferred orientation exhibits a complex dynamics. The orientation of topological rearrangements has an intrinsic bias and responds to cell shape changes with a delay. Motivated by these observations, we developed a continuum model that describes the rate of shear due to topological rearrangements by a linear constitutive equation that captures its behaviour.

In this work, we study the dynamical properties of this continuum model, including a response with memory of topological rearrangements to cell shape, as well as active contributions to stress and shear due to topological rearrangements.

In section \ref{sec:continuumTissue}, we present a generic hydrodynamic theory of flowing tissues with memory. We introduce hydrodynamic fields corresponding to the observable cell properties (Figure \ref{fig:cellularProcesses} C). Note that we use the word hydrodynamic here to denote a field relaxing slowly on large spatial scales. 

In Section \ref{sec:constitutiveRelatinos} we construct generic linear constitutive equations for a polar tissue, characterising the tissue stress as well as the shear created by topological rearrangements. A set of phenomenological coefficients is introduced in these equations, characterizing the response of the tissue. These coefficients are emerging properties of a tissue and can be experimentally measured. Similar to active gels \cite{Kruse2005, callan2011hydrodynamics}, an active stress can exist in the system from the forces generated inside cells by the cytoskeleton. In addition, we introduce an additional active term, distinct from an active stress, which describes active anisotropic topological rearrangements internally driven by the system. 

In Section \ref{sec:shearFlows} we consider an exponentially decaying memory kernel in the constitutive equation for shear, which was found to account for the behavior of the shear due to topological rearrangements in the {\it Drosophila} wing morphogenesis \cite{wingMorpho}. We explore the physics predicted by this model. In Sections  \ref{sec:tissueShearRheology} and \ref{sec:mechanicalNetork} we study the influence of memory effects on the tissue rheology. These memory effects can lead to oscillations in the tissue. We then show that the memory effects give rise to an effective inertia which is not related to the physical mass of the tissue. In Section \ref{sec:viscoelastNematic} we discuss the relation between the tissue model proposed here and the behaviour of an active viscoelastic nematic gel close to equilibrium. In Section \ref{sec:activityFlows} we discuss the dynamics of the tissue and cell deformation in a rectangular homogeneous tissue, and we find qualitative differences between tissues driven by the active stress and flows driven by active topological rearrangements, depending on boundary conditions. Finally, in Section \ref{sec:Couette} we discuss active tissue Couette flow where a stall force has to be applied to stop tissue from flowing spontaneusly.

\section{Hydrodynamic description of a flowing tissue}
\label{sec:continuumTissue}
\subsection{Cell density balance}
We consider a tissue consisting of cells with local cell number density $n$. The cell number can change through cellular events such as division and extrusions. Denoting the rates of cell divisions and extrusions per cell $k_d$ and $k_e$ respectively, the balance of cell number can be expressed as 
\begin{align}\label{eq:cellNumberBalance}
\partial_t n + \partial_k\left(n v_k\right) &= n \left(k_d - k_e\right) \quad ,
\end{align}
where we have introduced the cell velocity field $v_i$.
\subsection{Velocity gradient}
Deformations of the tissue result from spatial inhomogeneities of cell velocity $v_i$,  described by the velocity gradient tensor
\begin{align}
v_{ij} = \partial_j v_i \quad.
\end{align}
The velocity gradient matrix can be uniquely decomposed into a sum of isotropic, traceless symmetric and antisymmetric terms
\begin{equation}\label{eq:velocityGradientDecomposition}
v_{ij}=\frac{1}{d}v_{kk} \delta_{ij} +\tilde{v}_{ij} +\omega_{ij} \quad ,
\end{equation}
where $d$ is the number of dimensions. In the following we will discuss two-dimensional tissues, but extension to the $d= 3$ case is straightforward.
The trace of the velocity gradient $v_{kk}$ describes local changes in tissue area, the traceless symmetric part $\tilde{v}_{ij} = (v_{ij} + v_{ji})/2- \delta_{ij}v_{kk}/d$ is the pure shear rate and the antisymmetric part $\omega_{ij}=(v_{ij} - v_{ji})/2$ corresponds to local rotation of the tissue.
\subsection{Cell properties}\label{sec:cellProperties}
We propose a description of tissue flows at the scale larger than the typical cell size, but we retain information on the properties of cell shapes. The cell shape is characterised by the average cell area $a$ and the cell elongation nematic tensor $Q_{ij}$ (Fig. \ref{fig:cellularProcesses} C) describing the cell shape anisotropy. The average cell area of a cellular patch can be defined as $a = 1/n$, where $n$ is the cell number density. The cell elongation nematic has a magnitude $Q$ characterising the strength of cell elongation, and an angle $\varphi$ characterising the orientation of the elongation axis (see \ref{appendixNematics}). Different definitions of cell elongation based on cell outlines have been proposed \cite{wingMorpho, Guirao2015, TrianglePaper,  AigouyCell}. In the framework of the hydrodynamic theory we propose, we expect these various definitions to modify phenomenological coefficients but to leave the hydrodynamic equations unchanged. However, we impose that the definition of tensor $Q_{ij}$ shoud be such that homogeneous tissue deformation in the absence of topological rearrangements gives rise to the cell elongation change
\begin{align}
\frac{\mathrm{D} Q_{ij}}{\mathrm{D} t} &=  \tilde{v}_{ij}  \quad ,
\end{align}
where $\mathrm{D}/\mathrm{D}t$ is a corotational convected derivative (see \ref{app:corotationalDerivative}).

In addition, intracellular components can be distributed anisotropically inside the cell. Proteins of the planar cell polarity pathways, for instance, are known to be distributed across opposite cell edges \cite{GoodrichStrutt}. Here we take into account the cell polarisation by introducing a nematic tensor field $q_{ij}$ that describes the orientation of anisotropic structures in cells (Fig. \ref{fig:cellularProcesses} C). If the average cell polarity is characterised by a vector field $\mathbf{p}$, the corresponding nematic tensor is obtained from
\begin{align}
q_{ij}=p_i p_j -\frac{1}{2}p^2 \delta_{ij} \quad .
\end{align}
Cell polarity vectors and nematic tensors can be experimentally measured, for instance from the distribution of polarity proteins on cellular junctions \cite{AigouyCell, sagner2012establishment}. As for the cell elongation tensor, a particular choice of polarity definition will affect the phenomenological coefficients of linear hydrodynamic equations but not their general form. 
\subsection{Cellular contributions to tissue flows}
Isotropic and shear tissue flows can be decomposed in contributions reflecting changes of cellular properties. First, we note that Eq. (\ref{eq:cellNumberBalance}) can be rewritten as an equation for the isotropic flow in terms of the average cell area $a$ and cell division and extrusion rates
\begin{align}
\label{IsotropicShearDecomposition}
v_{kk} &= \frac{1}{a}\frac{\mathrm{d}a}{\mathrm{d}t} + k_d - k_e,
\end{align}
where $\mathrm{d}/\mathrm{d}t$ is the convected derivative (see \ref{app:corotationalDerivative}). Therefore, the relative change in tissue area is equal to the relative cell area change plus the relative change in cell number.
Anisotropic tissue deformation stems from two sources, (1) cellular deformations which can be captured by a change of cell elongation $Q_{ij}$, and (2) topological rearrangements which include $T_1$ transitions, cell divisions and cell death (see Fig. \ref{fig:cellularProcesses} B). Therefore, the tissue shear flow rate $\tilde{v}_{ij}$ can be decomposed according to the following equation
\begin{align}\label{eq:shearDecomposition}
\tilde{v}_{ij} &= \frac{\mathrm{D} Q_{ij}}{\mathrm{D}t} + R_{ij}
\end{align}
where $R_{ij}$ is a shear rate due to topological rearrangements. Note that collective correlated movements of cells can also contribute to $R_{ij}$ when spatial fluctuations of rotation and growth are correlated with cell elongation fluctuations  \cite{wingMorpho, TrianglePaper}. Here we will consider an effective constitutive equation for $R_{ij}$ that does not distinguish the different contributions arising from topological rearrangements or correlation contributions arising from coarse-graining.

\subsection{Force balance}
Viscous forces typically dominate over inertial contributions at the scale of cells and tissues. We therefore write force balance in the low Reynolds number limit 
\begin{equation}\label{ForceBalance}
\partial_i \sigma_{ij} + f_j^{\text{ext}}= 0 \quad .
\end{equation}
Here, $\sigma_{ij}$ is the tissue stress, and $\mathbf{f}^{\text{ext}}$ is an external force density acting on the tissue.

\section{Constitutive relations}
\label{sec:constitutiveRelatinos}
In Eqs. \ref{eq:shearDecomposition} and \ref{ForceBalance} we have introduced the tensorial quantities $R_{ij}$ and $\sigma_{ij}$, which are determined by tissue properties. We now propose constitutive relations for these quantities, using general assumptions about physical processes acting in the tissue. 
We distinguish in the constitutive equations ``passive''  terms, which tend to relax cell elongation, and ``active'' terms, which act as forcing terms driving cellular flow and deformation. Note that tissues function out of equilibrium and active processes in the cell can also in general contribute to passive terms: for instance, the effective cellular elasticity can depend on the activity of molecular motors in the cell. If cells are polarised, cell force generation can be anisotropic, giving rise to anisotropic active stress and anisotropic active topological rearrangements. Both effects are taken into account in our constitutive relations through the cell nematic polarity tensor $q_{ij}$ introduced in Section \ref{sec:cellProperties}.

\subsection{Tissue stress}\label{sec:tissueStress}
The total two-dimensional tissue stress can be decomposed in a pressure $P$ and shear stress $\tilde{\sigma}_{ij}$
\begin{align}
\sigma_{ij} &= -P \delta_{ij} + \tilde{\sigma}_{ij} \quad .
\end{align}
The anisotropic stress in the tissue depends on the cell elongation $Q_{ij}$ and cell polarity $q_{ij}$. We consider a linear elastic response of the tissue stress to tissue elongation described by $Q_{ij}$. In addition we introduce an active anisotropic stress capturing anisotropic force generation in the cell
\begin{align}\label{eq:stressConstitutiveFull}
\tilde{\sigma}_{ij}(t)&= \int\limits_{-\infty}^t \phi_K(t - t') Q_{ij}(t') dt'+\int\limits_{-\infty}^t \phi_\zeta(t-t') q_{ij}(t') dt' \quad .
\end{align}
where the memory kernels $\phi_K$ and $\phi_\zeta$ have units of a two-dimensional elastic modulus divided by a time. They characterise the response of anisotropic tissue stress to cell elongation $Q_{ij}$ and cell nematic polarity $q_{ij}$ (Fig. \ref{fig:cellularProcesses} D - top), respectively. In general, the memory kernels are fourth order tensors, however, here we consider the case when all anisotropies have been accounted for by $Q_{ij}$ and $q_{ij}$ and thus the memory kernels are isotropic.

For completeness, we express the isotropic stress in the tissue as a linear response to the natural strain $\ln{(a/a_0)}$ of cell area $a$
\begin{align}\label{eq:isotropicStressConstitutive}
P= - \overline{K}\ln{\left(\frac{a}{a_0}\right)} \quad , 
\end{align}
where $\overline{K}$ is the isotropic elastic modulus and $a_0$ is the cell area in a pressure-free tissue. In what follows, we focus on the role of anisotropic stress and cell elongation.

\subsection{Shear rate due to topological rearrangements}\label{sec:topoRearrangements}
Tissue rheology is governed by cell rearrangements described by the tensor $R_{ij}$. In the spirit of linear response theory, we express $R_{ij}$ in terms of other relevant nematic quantities, the average cell elongation $Q_{ij}$ and the internal cell anisotropy $q_{ij}$. Taking into account memory effects, we write to linear order 
\begin{align}\label{eq:R}
R_{ij}(t) &= \int\limits_{-\infty}^{t} \phi_\tau(t - t') Q_{ij}(t') dt'+ \int\limits_{-\infty}^{t}\phi_\lambda(t-t') q_{ij}(t') dt' \quad .
\end{align}
The memory kernels $\phi_\tau(t)$ and $\phi_\lambda(t)$ have units of inverse squared time  and characterise the response of shear produced by topological rearrangements to cell elongation and active cellular processes, respectively (Fig. \ref{fig:cellularProcesses} D - bottom). Eq. \ref{eq:R} corresponds to the underlying assumption that oriented, anisotropic topological rearrangements depend on the average cell elongation and the orientation of planar cell polarity. Note that similarly to the constitutive equation for the tissue stress Eqs. \ref{eq:stressConstitutiveFull} and \ref{eq:isotropicStressConstitutive}, the constitutive equation \ref{eq:R} and the memory kernels $\phi_\tau$ and $\phi_{\lambda}$ depend on the material or tissue considered. We expect that passive, apolar foams can be characterised by the function $\phi_\tau$, which describes how bubbles in the foam rearrange as a response to shear stress applied to the foam. In a biological process driven by anisotropic force generation on cell bonds, such as germ band elongation \cite{LecuitNature, CollinetLennLecuit2015}, we expect the function $\phi_{\lambda}$ to contain information characterising how internal tissue anisotropy results in oriented topological rearrangements. The functions $\phi_\tau$ and $\phi_{\lambda}$ therefore carry key information about cellular processes in the tissue, and can be measured experimentally in different morphogenetic tissues, in the same way that a tissue elastic modulus or viscosity can be measured in a rheological experiment.

\section{Shear flows}
\label{sec:shearFlows}
Eqs. \ref{IsotropicShearDecomposition} to \ref{eq:R} constitute a system of equations which can be solved for the velocity field of the tissue $\mathbf{v}$, the average cell area $a$ and the average cell elongation field $Q_{ij}$, once the rates of cell division and extrusion $k_d$ and $k_e$, the cell nematic polarity tensor $q_{ij}$, the phenomenological coefficients and memory kernels, and boundary conditions  are specified. In the next section, we discuss simple limits of the hydrodynamic theory of tissue flows we propose here.

\subsection{Tissue shear rheology} \label{sec:tissueShearRheology}
\begin{figure}[ht]
\centering
\includegraphics[width= .8\columnwidth]{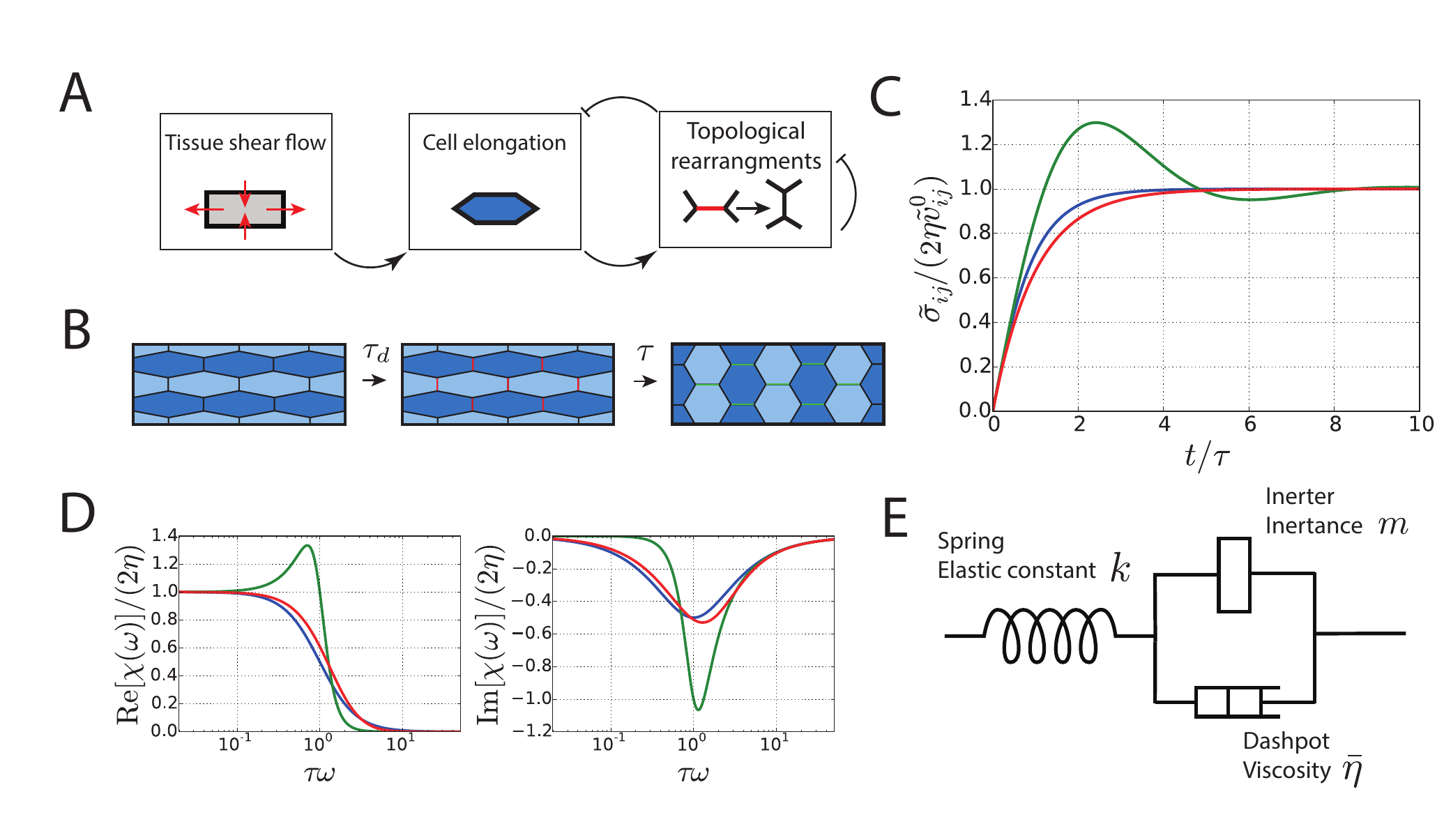}
\caption{ A) Schematics of the relation between cell elongation and topological rearrangements (Eqs. \ref{eq:shearDecomposition}, \ref{eq:R}). Shear due to topological rearrangements inhibits itself and cell elongation, which in turn induces shear due to topological rearrangements, forming a feedback loop with damping. Tissue shear flow can also modify cell elongation.   B) Schematics of the tissue response to a sudden change in cell elongation. The onset of cell elongation triggers topological rearrangements after a delay timescale $\tau_d$, which relax cell elongation over a timescale $\tau$. C) Shear stress response of a material subjected to a step in shear rate $\tilde{v}_{ij}$ (Eq. \ref{eq:stepResponse}). Blue line: for $\beta^2= -0.2 < 0$ the shear stress relaxes exponentially to a steady state value. Green line: for $\beta^2= 3 > 0$, the shear stress exhibits damped oscillations relaxing to the steady state value. Red line:  Case $\tau_d=0$, $\beta^2 = -1$, corresponding to a viscoelastic Maxwell material. D) Mechanical response function $\chi(\omega)$ for the three cases in C with corresponding line colors. E) A network containing a spring in series with a parallel connection of an inerter and a dashpot has equivalent rheology to a tissue with delayed topological rearrangements.
}
\label{fig:rheology}
\end{figure}

We first discuss the rheology of a homogeneous passive tissue subjected to external forces driving its deformation in the absence of active anisotropic cellular processes $q_{ij} = 0$. Moreover, we consider the case where the memory in the response of topological rearrangements to cell deformation arises from one dominating underlying relaxation process:  in that case the long time behavior of memory kernel $\phi_\tau$, introduced in Eq. \ref{eq:R}, is dominated by the largest timescale $\tau_d$
\begin{align}\label{eq:specificR}
R_{ij}(t) &= \frac{1}{\tau}\int\limits_{-\infty}^{t}\frac{1}{\tau_d} e^{-\frac{t - t'}{\tau_d}}Q_{ij}(t') dt'\quad .
\end{align}
which can also be written in a differential form 
\begin{align}\label{eq:diffR}
 \left(1 + \tau_d\partial_t\right) R_{ij} = \frac{1}{\tau} Q_{ij}\quad.
\end{align}
This equation, together with the shear flow decomposition Eq. \ref{eq:shearDecomposition}, describes the coupled dynamics of cell elongation and topological rearrangements, schematically represented in Fig. \ref{fig:rheology} A. Here, $\tau_d$ is the delay timescale over which cells integrate changes in cell elongation, such that changes in $R_{ij}$ are delayed by a time $\tau_d$ relative ti changes in cell elongation $Q_{ij}$. $\tau$ is a characteristic timescale of topological rearrangements, corresponding to the sensitivity of topological rearrangements to cell elongation (Figure \ref{fig:rheology} B). This form of response function was found to describe topological rearrangements during the morphogenesis of the {\it Drosophila} pupal wing \cite{wingMorpho}.  In the limit $\tau_d\rightarrow 0$, the tensor of topological rearrangements is simply proportional to the tensor of cell elongation $R_{ij} = Q_{ij}/\tau$.

Similarily, we discuss here a particular choice of shear stress constitutive relation in Eq. \ref{eq:stressConstitutiveFull} where memory effects relax on a timescale much shorter than other relevant timescales. Therefore
\begin{align}\label{eq:stressConstitutive}
\tilde{\sigma}_{ij} &= 2KQ_{ij} \quad ,
\end{align}
where $K$ is the anisotropic elastic modulus.

Assuming that topological rearrangements are described by Eq. \ref{eq:diffR}, a relationship between shear stress and shear flow can be derived by combining the shear flow decomposition Eq. \ref{eq:shearDecomposition}, the constitutive relation describing shear due to topological rearrangements Eq. \ref{eq:specificR} and the shear stress constitutive relation Eq. \ref{eq:stressConstitutive}
\begin{align}
\label{FlowStressRelation}
\tilde{v}_{ij}(t) &= \frac{1}{2K} \left(\partial_t\tilde{\sigma}_{ij} + \frac{1}{\tau}\int\limits_{-\infty}^{t}\frac{1}{\tau_d} e^{-(t - t')/\tau_d}\tilde{\sigma}_{ij}(t') dt'\right) \quad .
\end{align}
For simplicity we ignore here corotational terms. Eq. \ref{FlowStressRelation} can be rewritten in the frequency domain as
\begin{align}
\tilde{\sigma}_{ij}(\omega)&= \chi(\omega)\label{eq:freqRheology}
 \tilde{v}_{ij}(\omega) \\
\chi(\omega)&= 2\eta \frac{1 + i\tau_d\omega }{i\omega \tau (1 + i\tau_d\omega ) +1} \label{eq:freqResponse}
 \end{align}
 where $\eta=K\tau$ is a viscosity and $\tilde{v}_{ij}(\omega)$ and $\tilde{\sigma}_{ij}(\omega)$ are Fourier transforms of $\tilde{v}_{ij}(t)$ and $\tilde{\sigma}_{ij}(t)$, respectively (see \ref{app:fourier}).  In Eq. \ref{eq:freqRheology} we have introduced the frequency dependent mechanical response function $\chi(\omega)$, which characterises the rheology of the tissue. The function $\chi(\omega)$ is plotted on Fig. \ref{fig:rheology} D. The poles of $\chi(\omega)$ are in the upper half of complex plane, as required by causality.  We now discuss the form of the response function. 
 
For zero delay timescale, $\tau_d = 0$, the response function reduces to $\chi(\omega)=2\eta/(1+i\omega \tau)$, corresponding to a viscoelastic Maxwell material with relaxation time $\tau$ and long-time viscosity $\eta$. The viscoelastic behaviour can be understood as follows: on short timescale, tissue deformation results in cell elongation, and the emergence of an elastic stress in the tissue. On a timescale larger than $\tau$, cell elongation is relaxed by topological rearrangements, resulting in the relaxation of elastic stress and a fluid behaviour of the tissue.
 
For $\tau_d<\tau/4$, the poles have vanishing real parts and the system response is an exponential relaxation, similar to the case with zero delay timescale. 

For large enough delay time, $\tau_d>\tau/4$,  an original, oscillatory rheological behaviour arises. In that limit, the poles of the response function $\chi$ have non-zero real parts and the system exhibits damped oscillations. As a result, the coupled dynamics between cell elongation and topological rearrangements (Fig. \ref{fig:rheology} A) results in an oscillatory response of the tissue. To demonstrate this, we consider the stress response $\sigma_{ij}(t)$ to a step function in imposed shear $v_{ij}(t)=\tilde{v}_{ij}^0 \Theta(t)$, with $\Theta(t)$ the Heaviside step function. The resulting stress response $\tilde{\sigma}_{ij}(t)$ reads for $t>0$
\begin{align}\label{eq:stepResponse}
\tilde{\sigma}_{ij}(t)&= 2\eta\tilde{v}_{ij}^0 \left[1 - e^{-t/(2\tau_d)}\left(\frac{1-\beta^2}{2\beta}\sin{\left(\frac{t}{2\tau_d}\beta\right)} + \cos{\left(\frac{t}{2\tau_d}\beta\right)}\right)\right]\quad ,
\end{align}
where we have introduced $\beta = \sqrt{4\tau_d/\tau-1}$. For $\beta^2 < 0$ the stress relaxes exponentially, while for $\beta^2 > 0$ it exhibits damped oscillations  (Fig. \ref{fig:rheology} C).
Therefore, in experiment where a constant shear is imposed in the tissue should result in a transient oscillatory force response when the delay $\tau_d$ is sufficiently large.
\subsection{Representation by a simple mechanical network}
\label{sec:mechanicalNetork}
We now discuss whether the rheological properties of the material described by the response function $\chi(\omega)$ can be mapped to a simple rheological behaviour. We first note that the response function in Eq. \ref{eq:freqResponse} cannot be realised by any finite network of parallel and serial connections of springs and dashpots. This can be shown by inspecting the real and imaginary parts of $\chi(\omega)$
\begin{align}
\operatorname{Re}\chi(\omega) &= \frac{2\eta}{\tau^2 \omega^2 + (1 - \tau \tau_d \omega^2)^2} \\
\operatorname{Im}\chi(\omega) &= -2\eta\omega\frac{\tau -\tau_d + \tau\tau_d^2 \omega^2}{\tau^2 \omega^2 + (1 - \tau \tau_d \omega^2)^2} \quad .
\end{align}
We note that the imaginary part of the response function can become positive when $\tau_d > \tau$. As we now explain, a network of springs and dashposts in series and in parallel can only have a positive real part and a negative imaginary part of the response function. Let us consider two elements in series or in parallel in a rheological network, and assume that these two elements have the same sign of real and imaginary part of the response function at given frequency. One can verify then that the response function of the combined elements has the same sign of real and imaginary parts than the individual elements (\ref{sec:networkSpringDashpot}). The response function of a spring with elastic constant $k$ is $\chi_\text{spring} = - i k/\omega$, and the response function of a dashpot with viscosity $\overline{\eta}$ is $\chi_\text{dashpot} = \overline{\eta}$. As a result, any combination of spring and dashpot elements in series and in parallel  has a positive real part and a negative imaginary part.

However, it is easy to verify that the Laplace transform of the response function $\chi(t)$
\begin{align}
\widetilde{\chi}(s) &= 2\eta \frac{1 + \tau_d s}{\tau s\left(1 + \tau_d s\right) + 1}
\end{align}
is a positive real function. A positive real function is a rational complex function which is real for real values of $s$ and has a positive real part for  $\operatorname{Re}s > 0$. The Bott and Duffin synthesis theorem \cite{BottDuffin} for electrical circuits guarantees that a positive real response function can be reproduced by a network of resistors, capacitors and inductors. By drawing a mechanical analogy to electrical networks, one can verify that similarly, any rheological network with a positive real Laplace transform of the response function can be represented by a network of spring, dashpots, and inerters.  An inerter is an additional mechanical element corresponding to a capacitor in electrical networks, in the analogy where electrical current corresponds to stress and electric potential to shear rate \cite{SmithInerter}. The response function of an inerter is $\chi_\text{inerter} = i \omega m$ where $m$ is called inertance of the inerter and has units of mass for a two-dimensional system.

Interestingly, inerters are generally omitted from rheological schemes aimed at describing tissue rheology. Indeed, because biological tissues operate at low Reynolds numbers, inertial terms associated to the mass density of the tissue can be ignored when compared to viscous forces. We find here however that a delay in topological rearrangements introduces an effective inertial term, which does not come from physical masses in the system. We find that the response function of a tissue with delayed topological rearrangements is equivalent to a circuit made of a spring connected in series with a circuit of an inerter and dashpot connected in parallel (Fig. \ref{fig:rheology} E, \ref{sec:schematicResponse}). The inertance of the inerter is given by $m= 2 K \tau\tau_d$, the effective elastic constant of the spring is $k= 2K$ and the dashpot viscosity $\overline{\eta}=2\eta= 2K \tau$.   In Ref. \cite{wingMorpho}, we found that Eq. \ref{eq:freqRheology} accounted for experimental observations with $\tau\simeq 2\ h$ and $\tau_d\simeq 4\ h$. Assuming a typical tissue three-dimensional elastic modulus $K_{\text{3D}}\simeq 10\ \text{Pa}$ and characteristic cellular length-scale $l \simeq 10\mu m$, this corresponds to an inertance of $m= 2l K_{\text{3D}} \tau\tau_d\simeq 2\cdot 10^4$ kg. This inertance is very large compared to the actual physical mass of the system and arises from memory effects in the system.

\subsection{Viscoelastic nematic gel close to equilibrium}
\label{sec:viscoelastNematic}
We show in \ref{app:OnsagerTissue} that the shear decomposition equation \ref{eq:shearDecomposition} also applies to nematic viscoelastic gels. In a gel, the cell elongation tensor $Q_{ij}$ corresponds to the local elastic shear strain. In \ref{app:OnsagerTissue}, we derive for comparison constitutive equations for an active viscoelastic nematic gel close to equilibrium, where Onsager symmetry relations impose relations between the phenomenological coefficients relating fields and their conjugated thermodynamic forces \cite{de2013non}. Taking into account the internal dynamics of the nematic field in the gel, we find an effective constitutive equation for the tensor $R_{ij}$, which responds with a delay to changes in the local elastic shear strain, and contains an additional term involving a time derivative of elastic shear strain $Q_{ij}$ compared to Eq. \ref{eq:diffR}:
\begin{align}\label{eq:diffR_modified}
\left(1 + \tau_d\frac{\mathrm{D}}{\mathrm{D}t}\right) R_{ij} = \frac{1}{\tau} Q_{ij}+\alpha \frac{\mathrm{D} Q_{ij}}{\mathrm{D}t} \quad. 
\end{align}
In this work, we consider the simple case where $\alpha=0$. We discuss in \ref{app:OnsagerTissue} parameter regimes of the viscoelastic nematic gel theory where $\alpha$ is negligible.

\subsection{Activity-induced shear flows}\label{sec:activityFlows}
We now discuss the effects of the two distinct active processes introduced in Eqs. \ref{eq:stressConstitutiveFull} and \ref{eq:R}. We start from the constitutive equation for the tensor of topological rearrangements, Eq. \ref{eq:R}, and assume that the two memory kernels $\phi$ and $\phi_{\lambda}$ are exponential with the same relaxation timescale $\tau_d$. This leads to a differential equation for $R_{ij}$
\begin{align}\label{eq:fullR}
  \left(1 + \tau_d \partial_t\right)R_{ij} &= \frac{1}{\tau}Q_{ij} + \lambda q_{ij} \quad ,
\end{align}
where $\lambda$ characterises the magnitude of shear flow due to active oriented topological rearrangements. Moreover, we assume that the response of tissue shear stress to tissue polarity is instantaneous and has magnitude $\zeta$
\begin{align}\label{eq:fullStressConstitutive}
\tilde{\sigma}_{ij} &= 2K Q_{ij} + \zeta q_{ij} \quad .
\end{align}
\begin{figure}[h!]
\centering
 \includegraphics[width= .7\columnwidth]{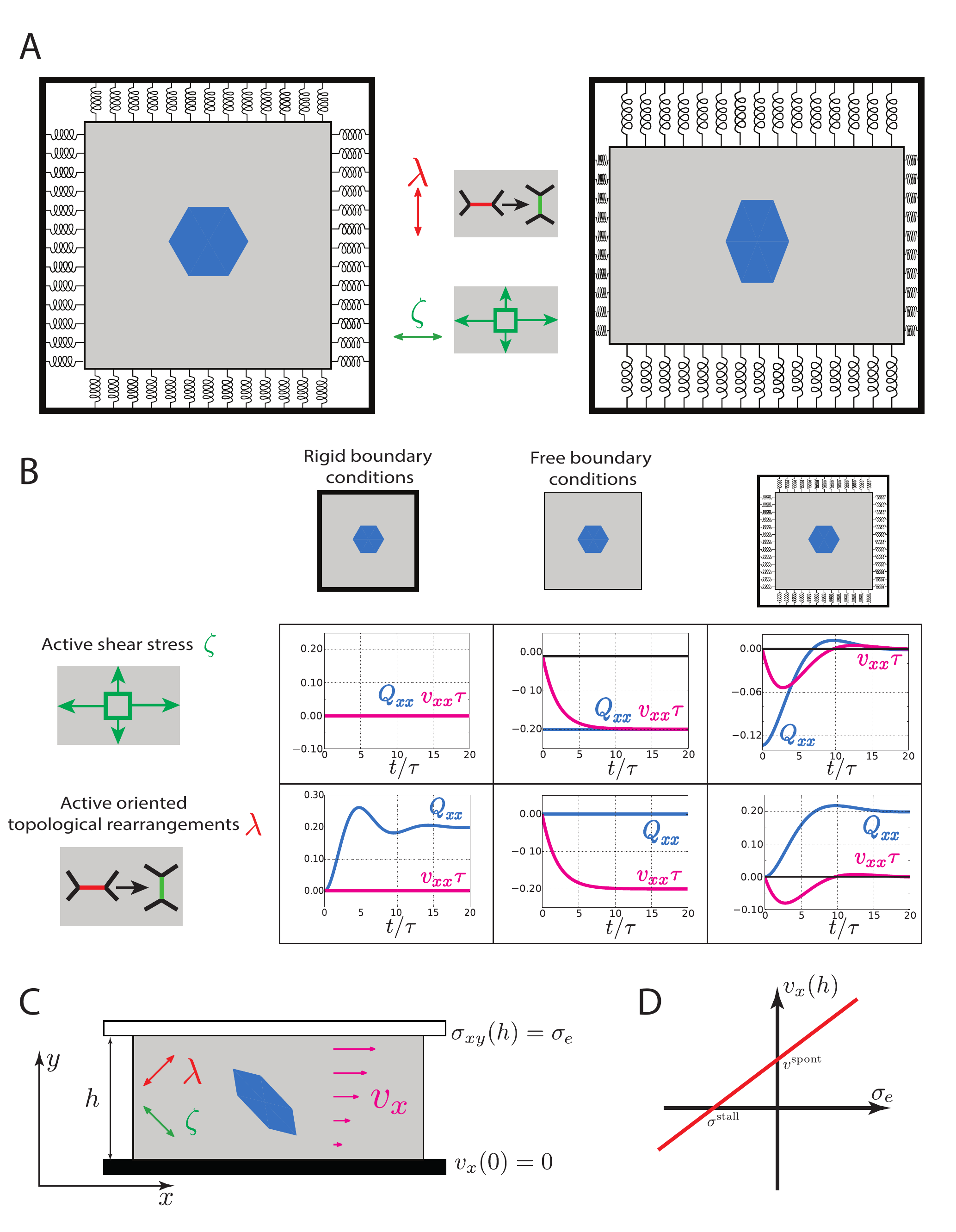} 
\caption{ A) A rectangular, homogeneous tissue, is attached to an external solid frame by springs. Active processes drive internal anisotropic tension and oriented topological rearrangements in the tissue. B) Time evolution of the shear and cell elongation of a rectangular tissue, subjected either to active stress ($\zeta\neq 0$ and  $\lambda = 0$, top row), or to active shear due to topological rearrangements ( $\zeta = 0$ and $\lambda \neq 0$, bottom row), under different boundary conditions. The active stress and active shear due to topological rearrangements give rise to different behaviour of cell elongation and tissue flow. The active shear coefficient $\gamma$ is set to be the same in all cases: $\gamma \tau = -\zeta /(2K) + \lambda\tau= -0.2$. Other parameters: $\tau_d/\tau=2$ and in the third column, the spring constants are $k_x/K=k_y/K=1$. Note that a Dirac delta peak at $t=0$ in $v_{xx}(t)$ is not plotted. C) Active Couette flow, reminiscent of tissue flow in developing Drosophila genital imaginal disc \cite{sato2015left}. The tissue is constrained to have width $h$ and the bottom of the tissue is fixed $v_x(0)= 0$ while the top tissue margin is free to slide $\sigma_{xy}(h)= 0$. D. Relation between the velocity of the upper interface and the shear stress applied. The tissue has a spontaneous flowing velocity $v^{\rm{spont}}$ and a stalling stress $\sigma^{\rm{stall}}$.
}
\label{fig:activePolarProcesses}
\end{figure}
Let us first discuss a simple but instructive case of a convergent extension process in a stress free, homogeneous, polarised tissue. Two distinct active processes can drive tissue and cell deformation: active topological rearrangements, characterised by the coefficient $\lambda$ in Eq. \ref{eq:fullR}, and active stress generation, characterised by the coefficient $\zeta$ in Eq. \ref{eq:fullStressConstitutive}. At steady state, using Eqs. \ref{eq:shearDecomposition}, \ref{eq:fullR} and \ref{eq:fullStressConstitutive}, one finds that as a result of these active processes in the tissue, the tissue deforms with the shear rate
\begin{align}\label{eq:freeTissueFlow}
\tilde{v}_{ij} &= \left(-\frac{\zeta }{2K \tau} + \lambda\right) q_{ij} \quad .
\end{align}
From Eq. \ref{eq:freeTissueFlow}, one finds that at long timescales, the tissue is constantly deforming, with a steady-state shear flow controlled by the active shear coefficient $\gamma= -\zeta/(2\eta) + \lambda$. Therefore for $\gamma \neq 0$, non-zero flows are present in the steady state. 
Note that the contributions of the two active processes cannot be distinguished by observing only the tissue flow for a stress free tissue. However, observing the cell elongation together with the flow allows to identify the active shear stress $\zeta q_{ij}= - 2K Q_{ij}$ and thus to distinguish the two active processes.

We now discuss the behaviour of our model in a tissue subjected to different combinations of active processes and boundary conditions. We consider a rectangular shaped, homogeneous tissue, oriented such that polarity axis is along the $x$ axis: $q_{xx} = 1, q_{xy} = 0$ (see Fig. \ref{fig:activePolarProcesses}A). The tissue is constrained in space by an elastic material connected to a solid frame. The elastic material provides resistance to changes in tissue length and height. We define the natural strain variables $L = \ln{(l/l_0)}$ and $H = \ln{(h/h_0)}$, where $l$, $h$ are tissue length and tissue width. Introducing the elastic moduli of the surrounding material $k_x$ and $k_y$, the external stress acting on the tissue can be written
\begin{align} \label{eq:boundaryRectangle}
\begin{split}
\sigma_{xx} &= -k_x L\\
\sigma_{yy} &= -k_y H \quad .
\end{split}
\end{align}
We assume that for $t<0$, the tissue is at rest, not subjected to active anisotropic processes, that cells are isotropic and that there is no stress in the elastic material surrounding the tissue. We then assume that active anisotropic processes are turned on at $t=0$ and are constant for $t>0$. We then solve for the tissue shape and cell elongation using Eq. (\ref{eq:fullR}). For simplicity, we discuss here only the case $k_x = k_y=k$. In this case, the tissue area $A = A_0\exp{(H(t) + L(t))}$  is conserved. The general solution is given in \ref{appendixConvergentExtension}. We find for $t\geqslant0$;
\begin{align}
\label{eq:exampleSolutionL}
L(t) &= \gamma\frac{2\tau}{\nu}\left[1 - \left(1 + \frac{\zeta}{2 \tau K \gamma}\frac{\nu}{\mu}\right)e^{-t/(2\tau_d)}\left(\frac{1}{s}\sin{\frac{s}{2\tau_d}t} + \cos{\frac{s}{2\tau_d}t}\right)\right] \\
 H(t)  &= -L(t), \label{eq:LHrelation}\\ 
v_{xx}(t) &= 4 \frac{\gamma}{s \mu}\left(1 + \frac{\zeta}{2K \tau \gamma}\frac{\nu}{\mu}\right) e^{-t/(2\tau_d)}\sin{\frac{s}{2\tau_d}t} - \frac{\zeta}{\mu K}\delta(t)\\\label{eq:exampleSolutionQ}
Q_{xx}(t) &= \frac{1}{2K}\left(-kL(t) - \zeta\right)
\quad ,
\end{align}
where
\begin{align}
s &= \sqrt{4\frac{\tau_d}{\tau}\frac{\nu}{\mu} - 1}\\
\mu &= 2\left(1 + \frac{k}{2K}\right) \\
\nu &= \frac{k}{K} \quad .
\end{align}
and the corresponding dynamics of the shear rate $v_{xx}$ and cell elongation $Q_{xx}$ are plotted on Fig. \ref{fig:activePolarProcesses} B for a few parameter values. We distinguish an oscillatory solution when $s^2 > 0$ corresponding to high values of $\tau_d$  and non-oscillatory dynamics when $s^2 < 0$ corresponding to low values of $\tau_d$. Note that for $s^2 < 0$, Eq. \ref{eq:exampleSolutionL} can be written in terms of hyperbolic functions (see \ref{appendixConvergentExtension}). When boundary conditions are stress free, $\nu/\mu= 0$, $s^2= -1$ and the solution cannot be oscillatory. When imposing rigid boundary conditions, $\nu/\mu = 1$ and the parameter $s$ is equal to the parameter $\beta$ in Eq. \ref{eq:stepResponse}. In general these two coefficients are related by $ (s^2+1) \mu/\nu= \beta^2 + 1$. The factor $\nu/\mu$ is related to boundary conditions; by imposing firmer boundary conditions, this factor is increased. Therefore, the oscillatory behavior disappears for decreased rigidity of the boundary springs. However, oscillatory flow does not necessarily appear at high external rigidity, due to the constraint $\nu/\mu < 1$. 

Using the solution in Eqs. \ref{eq:exampleSolutionL} and \ref{eq:exampleSolutionQ}, we now discuss the cases of free, rigid, and intermediate boundary conditions, and consider the difference between flows and cell elongation induced by  the anisotropic active stress $\zeta q_{ij}$ and anisotropic topological rearrangements $\lambda q_{ij}$ (Fig. \ref{fig:activePolarProcesses} A). We take here $\lambda<0$ and $\zeta>0$, such that the active stress is larger along the $x$ direction and horizontal bonds are actively removed. Tissue flow and cell elongation depend crucially on boundary conditions (Fig. \ref{fig:activePolarProcesses}B).

For rigid boundary conditions, the tissue can not deform and the active anisotropic stress has no effect. Active topological rearrangements however drive cell elongation along the $x$ direction. The process reaches steady-state when topological rearrangements driven by cell elongation are balanced by active topological rearrangements, and a final cell elongation $Q_{xx}= -\lambda \tau$ is reached. 

For free boundary conditions, active anisotropic stresses result in cell elongation along the $y$ direction and tissue flow. In the steady state limit, the cell elongation reaches the value $Q_{xx}= -\zeta/(2K)$. Active topological rearrangements do not result in cell deformation, but generate tissue shear.

For intermediate values of boundary spring elasticity, the tissue flows until the boundary springs deform sufficiently to balance stresses in the tissue. When the tissue shape  reaches steady state, flows vanish and the average cell elongation reaches the same value as in the case of rigid boundary conditions, $Q_{xx}= -\lambda \tau$. 

In the cases described above, when boundary conditions are such that the tissue eventually stops flowing, a steady state is reached only when the shear created by topological rearrangements also vanishes. This occurs when topological rearrangements  induced by cell elongation  and spontaneous, active topological rearrangements  balance each other. This process selects a value of the cell elongation tensor, $Q_{ij}=-\lambda\tau q_{ij}$ (see Eq. \ref{eq:fullR}). Therefore, by controlling active topological rearrangements, a tissue should be able to establish a fixed value of cell elongation at steady-state.

In addition, we note that although both active shear stress and active topological rearrangements can induce tissue flows (Fig. \ref{fig:activePolarProcesses} B), these two processes affect cell shape differently. This suggests that the relative contribution in tissue morphogenesis of these two active processes can be distinguished by observing both cell and tissue shape changes.

\subsection{Active tissue Couette flow}
\label{sec:Couette}
We now discuss the Couette flow of a tissue confined between two parallel plates. We consider a two-dimensional tissue which is constrained in the direction $y$ to have fixed width $h$, it is translationally invariant in the other direction and is firmly fixed at the bottom side of tissue $y= 0$ (see Fig. \ref{fig:activePolarProcesses} C).  Cell polarity is assumed to be homogeneous in space. We assume for simplicity that the tissue is incompressible, $v_y= 0$, such that the only non-vanishing component of the shear flow is $\tilde{v}_{xy}= \partial_yv_x/2$. On the top side, a fixed external shear stress $\sigma_{xy}(y=h)=\sigma_e$ is imposed on the tissue. The tissue is immobile for $t<0$, and active processes are turned on at $t=0$ and are then constant in time. The general solution is given in \ref{app:Couette}. 

We first consider the case where the top side of the tissue $y= h$ is free to slide, $\sigma_e= 0$. 
The Couette flow velocity increases and reaches a steady-state over the timescale $\tau_d$
\begin{align}\label{eq:CouetteMain}
\partial_yv_x(t)&= 2\left(\lambda - \frac{\zeta}{2K\tau}\right)q_{xy}\left(1 - e^{-\frac{t}{\tau_d}}\right) \quad .
\end{align}
Similar to earlier examples, this flow can be produced by either active stress and active shear due to topological rearrangements. Measuring the cell elongation component in the steady state $Q_{xy}= -\zeta/(2K) q_{xy}$ allows to determine how much each of the two active processes contributes to the tissue flow. If the component $q_{xx}$ of the cell polarity is present it does not contribute to the tissue flows but determines the steady state value of cell elongation component $Q_{xx}= - \tau\lambda q_{xx}$ since the boundary conditions in this direction are rigid. This situation is reminiscent of flows observed in the Drosophila genitalia rotation and vertex model simulations intended to reproduce this process \cite{sato2015left, sato2015cell}.

Finally, we obtain for arbitrary external shear stress $\sigma_e$ at steady-state
\begin{align}
v_{x}(h) &=  \frac{h}{K\tau}\sigma_e + 2h\left(\lambda - \frac{\zeta}{2 K \tau}\right)q_{xy},
\end{align}
which is a relation between the velocity of the upper interface, the external shear applied to the tissue and the spontaneous flowing velocity of the tissue $v^{\rm{spont}}=2h(\lambda-\zeta/(2K\tau))$. The effective friction coefficient of the tissue layer is $K\tau/h$, while a stalling stress $\sigma_e=\sigma^{\rm{stall}}=(\zeta/2-K\tau\lambda)q_{xy}$ must be applied to stop the tissue from spontaneously flowing (Figure \ref{fig:activePolarProcesses} D). It would be interesting to see if rheological experiments can allow to measure these quantities.

\section{Discussion}
We have presented a hydrodynamic theory of tissue shear flows which explicitly accounts for topological rearrangements at the cellular scale. We have introduced active terms influencing the shear stress as well as oriented topological rearrangements of cells, coupled to the nematic field $q_{ij}$. The theory applies to both effectively two-dimensional epithelia, which have been most widely studied \cite{LecuitLenne2007, Keller2012, HeisenbergBellaiche2013}, and to three-dimensional tissues, which can also undergo topological rearrangements \cite{warga1990cell, keller2008forces}. We have introduced phenomenological parameters characterising the response of shear stress to cell shape anisotropy and cell polarisation ($\phi_K$ and $\phi_\zeta$ in Eq. \ref{eq:stressConstitutiveFull}) and shear due to topological rearrangements  ($\phi_\tau$ and $\phi_{\lambda}$ in Eq. \ref{eq:R}). These parameters can be experimentally measured, similarly to an elastic modulus or a viscosity, and we expect that future analysis of morphogenetic processes will involve the quantification of their values.

We find that the dependency of topological rearrangements on cell shape generically leads to tissue fluidification. Without delay between cell elongation and topological rearrangement, the long timescale tissue viscosity depends on cellular elasticity and on a characteristic timescale of topological rearrangements (Eq. \ref{eq:freqResponse}) .  

We have also introduced a delay in the response of topological rearrangements to cell shape change (Eq. \ref{eq:specificR}) through an exponential memory kernel, motivated by experimental observations in the {\it Drosophila} wing epithelium during pupal morphogenesis \cite{wingMorpho}. For delay timescale sufficiently smaller than the characteristic timescale of topological rearrangements, $\tau_d< \tau/4$, the qualitative behavior of the tissue is not different from the one without delay in the response. However, sufficiently large delay timescale $\tau_d > \tau/4$ leads to a novel rheological behavior, with an oscillatory response of the tissue to the imposed shear. The response of the tissue can be described by a simple rheological scheme, involving a spring, a dashpot, and an effective inertial element (Fig. \ref{fig:rheology} E). Such an effective inertial response is not generally taken into account as tissues operate at low Reynolds number, but we show that complex tissue time-dependent behaviour can indeed show an effective inertial behaviour, unrelated to the tissue's real mass. Recent works have proposed alternative mechanisms giving rise to effective inertial behaviour, possibly arising from the dynamics of cell self-propelling forces \cite{serra2012mechanical} or concentration and polarity fields coupled to tissue flow and internal stresses \cite{Banerjee2015stressWaves, notbohm2016cellular}.

The theory we propose here also accounts for autonomously produced stress and oriented topological rearrangements that are a consequence of active processes in the tissue. They drive tissue flows and cell elongation changes on long timescales even in the absence of external forces. In general, active stress and active topological rearrangements may occur together in tissue morphogenesis, and both drive tissue and cell shape changes. Quantification methods must be developed to characterise the effects of these active processes \cite{wingMorpho}. 

Our theory makes a number of experimentaly testable predictions. Most notably, perturbing boundary conditions around a developing homogeneous tissue affects significantly the dynamics of both tissue flow and cell elongation. We predict that the topological rearrangements will drive flow when tissue is free to deform, but will drive instead cell elongation changes in the direction perpendicular to the original shear flow when the tissue is prevented from deforming. Intuitively, active topological rearrangement force cells to change their neighbour relationships, driving cell elongation; when the tissue is free, cell elongation can relax, driving tissue deformation; while when tissue deformation is prevented, cell elongation is maintained.  In contrast, active stress can also generate a flow in a free tissue but will not drive cell elongation changes when tissue deformation is stopped. Possibly, experiments aiming at preventing tissue flow could allow to distinguish between these contributions.

We have focused here on anisotropic flows. We expect that future work combining a description of isotropic and anisotropic flows in tissues, incorporating the effect of cell division and cell extrusion, will allow a full picture of tissue morphogenesis to be obtained. 

Finally, we have described here the case of a flat tissue. It would be interesting to explore how the theory proposed here extends to the more general case of a curved tissue.

\section*{Acknowledgements}
MP, AN, MM, FJ and GS acknowledge funding by the Max-Planck-Gesellschaft. RE acknowledges a Marie Curie fellowship from the EU 7th Framework Programme (FP7). SE acknowledge funding from the ERC. MP, MM, SE and FJ acknowledge funding by Bundesministerium f{\"u}r Bildung und Forschung. MM also acknowledges funding from the Alfred P. Sloan Foundation, the Gordon and Betty Moore Foundation, and NSF-DMR-1352184. GS acknowledges support by the Francis Crick Institute which receives its core funding from Cancer Research UK (FC001317), the UK Medical Research Council (FC001317), and the Wellcome Trust (FC001317).

\appendix

\section{Nematic tensors in 2D}\label{appendixNematics}
The nematic tensor $Q_{ij}$ can be expressed by its cartesian components in two dimensions
\begin{align}
\begin{pmatrix}
Q_{xx} & Q_{xy}\\
Q_{xy} & -Q_{xx} 
\end{pmatrix}\quad ,
\end{align}
It can also be rewritten using polar components
\begin{align}
\begin{pmatrix}
Q\cos{2\varphi} & Q\sin{2\varphi}\\
Q\sin{2\varphi} & -Q\cos{2\varphi}
\end{pmatrix} \quad ,
\end{align}
where $Q$ and $\varphi$ are respectively the magnitude and angle of the nematic relative to the $x$ axis.

\section{Conventions}
\subsection{Convected and corotational derivatives}\label{app:corotationalDerivative}
The convected derivative of a scalar field $S$ is defined as 
\begin{align}
\frac{\mathrm{d}S}{\mathrm{d}t} &= \frac{\partial S}{\partial t} + v_k \partial_kS \quad .
\end{align}
The convected corotational derivative of a vector field $U_i$ is defined as
\begin{align}
\frac{\mathrm{D} U_i}{\mathrm{D} t} &= \frac{\partial U_i}{\partial t} + v_k \partial_k U_i + \omega_{ij}U_j \quad ,
\end{align}
and for a tensor field $V_{ij}$ 
\begin{align}
\frac{\mathrm{D} V_{ij}}{\mathrm{D} t} &= \frac{\partial V_{ij}}{\partial t} + v_k \partial_k V_{ij} + \omega_{ik}V_{kj}+ \omega_{jk}V_{ik} \quad .
\end{align}
\subsection{Fourier transform}\label{app:fourier}
The following convention for Fourier transform is used:
\begin{align}
f(\omega) &= \int\limits_{-\infty}^\infty f(t) e^{-i\omega t}dt\\
f(t) &= \frac{1}{2\pi}\int\limits_{-\infty}^\infty f(\omega) e^{i\omega t} d\omega \quad.
\end{align}
Here $f(\omega)$ is the Fourier transform of the function $f(t)$.

\section{Rheological networks}

\subsection{Real and imaginary parts of a rheological network response function}\label{sec:networkSpringDashpot}
We discuss here some properties of the response function $\chi(\omega)$ of the serial and parallel connections of two mechanical elements with response functions $\chi_1(\omega)$ and $\chi_2(\omega)$.
The mechanical response function of a parallel connection of these elements is 
\begin{align}\label{eq:parallelConnection}
\chi_\text{parallel} = \chi_1 + \chi_2
\end{align}
so that 
\begin{align}
\label{eq:parallelRe}
\operatorname{Re} \chi_{\text{parallel}} &= \operatorname{Re} \chi_1 + \operatorname{Re} \chi_2\\
\label{eq:parallelIm}
\operatorname{Im} \chi_{\text{parallel}} &= \operatorname{Im} \chi_1 + \operatorname{Im} \chi_2 \quad .
\end{align}
The response function of a serial connection is 
\begin{align}\label{eq:serialConnection}
\chi_{\text{serial}}= \frac{\chi_1 \chi_2}{\chi_1 + \chi_2} \quad, 
\end{align}
which can be written in terms of real and imaginary parts 
\begin{align}\label{eq:serialRe}
\operatorname{Re}\chi_{\text{serial}} &= \frac{\operatorname{Re} \chi_1 \operatorname{Re} \chi_2\left(\operatorname{Re} \chi_1 + \operatorname{Re} \chi_2\right) + \operatorname{Re} \chi_1 \left[\operatorname{Im} \chi_2\right]^2 + \operatorname{Re} \chi_2 \left[\operatorname{Im} \chi_1\right]^2 }{\left(\operatorname{Re} \chi_1 + \operatorname{Re} \chi_2\right)^2 + \left(\operatorname{Im} \chi_1 + \operatorname{Im} \chi_2\right)^2}\\
\label{eq:serialIm}
\operatorname{Im}\chi_{\text{serial}} &= \frac{\operatorname{Im} \chi_1 \operatorname{Im} \chi_2\left(\operatorname{Im} \chi_1 + \operatorname{Im} \chi_2\right) + \operatorname{Im} \chi_1 \left[\operatorname{Re} \chi_2\right]^2 + \operatorname{Im} \chi_2 \left[\operatorname{Re} \chi_1\right]^2 }{\left(\operatorname{Re} \chi_1 + \operatorname{Re} \chi_2\right)^2 + \left(\operatorname{Im} \chi_1 + \operatorname{Im} \chi_2\right)^2}
\end{align}
By inspecting Eqs. \ref{eq:parallelRe} and \ref{eq:serialRe} we conclude that if the real parts of the response functions of two elements in a serial or parallel connection have the same sign, this sign is preserved in the real part of the response function of the connection. The same is true for the imaginary parts (Eqs. \ref{eq:parallelIm} and \ref{eq:serialIm}).

\subsection{Response function of a rheological network with a spring in series with a parallel connection of an inerter and a dashpot}\label{sec:schematicResponse}
We calculate here the response function of the mechanical network shown in Fig. \ref{fig:rheology}E. Using Eq. \ref{eq:parallelConnection} we find that the response function of a parallel connection of an inerter with inertance $m$ and dashpot with viscosity $\overline{\eta}$ reads
\begin{align}\label{eq:mEta}
\chi_{m, \overline{\eta}}= i \omega m + \overline{\eta} \quad .
\end{align}
The full response function is found using Eq. \ref{eq:serialConnection} for a serial connection of a spring with elastic constant $k$ and parallel connection of an inerter and dashpot
\begin{align}
\chi&= \frac{\frac{k}{i\omega}\left(i\omega m +\overline{ \eta}\right)}{i\omega m + \overline{\eta} + \frac{k}{i\omega}}\nonumber\\
  &= \overline{\eta} \frac{1 + i\omega \frac{m}{\overline{\eta}}}{i\omega \frac{\overline{\eta}}{k}\left(1 + i\omega \frac{m}{\overline{\eta}}\right) + 1} \quad .
\end{align}
Comparing with the Eq. \ref{eq:freqResponse} we can identify $\overline{\eta} = 2K\tau$, $k= 2K$ and $m= 2K\tau\tau_d$.

\section{Viscoelastic nematic gel close to equilibrium}\label{app:OnsagerTissue}
Here we derive hydrodynamic equations for a viscoelastic nematic gel close to equilibrium and we discuss parameter regimes which reproduce Eq. \ref{eq:diffR}. Our derivation is similar to the one given in the Ref. \cite{callan2011hydrodynamics}, but for a one-component, nematic gel. We start by writing conservation equations for density and momentum
\begin{align}
\partial_t \rho + \partial_i\left(\rho v_i \right) &= 0 \\
\partial_t g_i - \partial_j \sigma_{ij}^{\mathrm{tot}} &= 0
\end{align}
where $g_i = \rho v_i$ and $\sigma_{ij}^{\mathrm{tot}}$ is the total two-dimensional stress. We treat the local elastic shear strain $Q_{ij}$ as a thermodynamic state variable. We also include in our description a nematic order parameter $q_{ij}$ describing other internal anisotropies, motivated by the cell polarity in tissues. Here we will discuss the case when $q_{ij}$ is not spontaneously generated in equilibrium. Assuming that the dynamics of elastic shear strain $Q_{ij}$ and nematic order parameter $q_{ij}$ are sufficiently slow, we introduce the free energy density of the gel as $f(\rho, Q_{ij}, q_{ij})$. 

We define the elastic shear stress as being the thermodynamic conjugate quantity to the elastic shear strain $Q_{ij}$
\begin{align}
\sigma_{ij}^{\mathrm{el}} &= \frac{\delta f}{\delta Q_{ij}} \quad .
\end{align}

We also allow for ATP driven active processes described by the chemical reaction rate $r$ and the chemical potential difference between ATP molecule and hydrolysis products $\Delta \mu$. 

Using the free energy density $f(\rho, Q_{ij}, q_{ij})$, we find the expression for the density of entropy production rate $\dot{\sigma}$
\begin{align}
T \dot{\sigma}&= \sigma_{ij} v_{ij}  - \sigma_{ij}^\mathrm{el} \frac{\mathrm{D}Q_{ij}}{\mathrm{D}t} + H_{ij}  \frac{\mathrm{D}q_{ij}}{\mathrm{D}t} + r \Delta \mu \quad.
\end{align}
Here $\sigma_{ij}= \sigma_{ij}^{\mathrm{tot}} + \rho v_i v_j + P \delta_{ij}$, $v_{ij} = 1/2(\partial_i v_j + \partial_j v_i)$ is the symmetric part of velocity gradient tensor and $H_{ij}$ is the nematic molecular field conjugate to $q_{ij}$ defined as
\begin{align}
H_{ij} &= -\frac{\delta f}{\delta q_{ij}}\quad.
\end{align}

For simplicity, we ignored here terms associated to the gradients of chemical potential and elastic shear strain, and we discuss traceless symmetric components. Identifying the thermodynamic fluxes $\tilde{\sigma}_{ij}$, $\mathrm{D}Q_{ij}/\mathrm{D}t$, $\mathrm{D}q_{ij}/\mathrm{D}t$ and $r$, and the corresponding forces $\tilde{v}_{ij}$, $-\sigma_{ij}^\mathrm{el}$, $H_{ij}$ and $\Delta\mu$, we write the phenomenological equations
\begin{align}
\tilde{\sigma}_{ij} &= 2 \eta \tilde{v}_{ij} +\sigma_{ij}^\mathrm{el} - \beta_1 H_{ij} + \left(\Theta q_{ij} + \overline{\Theta}  Q_{ij}\right) \Delta\mu\\
\frac{\mathrm{D} Q_{ij}}{\mathrm{D}t} &= \tilde{v}_{ij} - \Gamma \sigma_{ij}^\mathrm{el} + \frac{1}{\beta_2} H_{ij}+ \left(\psi q_{ij} + \overline{\psi}Q_{ij} \right)\Delta\mu \label{eq:QDynamics}\\
\frac{\mathrm{D} q_{ij}}{\mathrm{D}t} &= \beta_1\tilde{v}_{ij} - \frac{1}{\beta_2}\sigma_{ij}^\mathrm{el} + \gamma H_{ij} + \left(\theta q_{ij}  + \overline{\theta} Q_{ij}\right) \Delta\mu\label{eq:qDynamics}\\ 
\begin{split}
r &=   -\left(\Theta q_{ij} + \overline{\Theta} Q_{ij}\right)\tilde{v}_{ij} - \left(\psi q_{ij} + \overline{\psi} Q_{ij}\right)\sigma_{ij}^\mathrm{el} \\
&+ \left(\theta q_{ij} + \overline{\theta} Q_{ij}\right)H_{ij}+ \Lambda\Delta \mu\quad .
\end{split}
\end{align}
where Onsager symmetry relations have been taken into account. We have included ATP consumption terms coupling to fluxes of different tensorial order through Onsager coefficients proportional to the elastic shear strain $Q_{ij}$ and nematic order parameter $q_{ij}$. Note that since elastic shear strain  is conjugate to the elastic shear stress, the Onsager coefficient relating $\tilde{v}_{ij}$ to the change of shear strain $Q_{ij}$ can be set to 1 without loss of generality, see \cite{callan2011hydrodynamics}. 

One can already note that Eq. \ref{eq:QDynamics} is in the form of Eq. \ref{eq:shearDecomposition} if we identify $R_{ij} = \Gamma \sigma_{ij}^{\mathrm{el}} -  H_{ij}/\beta_2 - \psi \Delta \mu q_{ij} - \overline{\psi}\Delta\mu Q_{ij}$. We now show that this general description can result in the delayed response of $R_{ij}$ discussed in the main text, produced by the relaxation dynamics of $q_{ij}$. 
%

\subsection{Passive gel}\label{app:OnsagerPassive}
Let us first consider the case of a passive gel which would not consume ATP.  Using Eqs. \ref{eq:QDynamics} and \ref{eq:qDynamics}, the tensor $R_{ij}$ can be expressed as a function of the shear strain $Q_{ij}$
\begin{align}\label{eq:passiveTissue}
\left[1 + \frac{1}{\gamma\kappa\left(1 - \frac{\beta_1}{\gamma\beta_2}\right)} \frac{\mathrm{D}}{\mathrm{D}t}\right] R_{ij} &= 2K\frac{\Gamma - \frac{1}{\beta_2^2 \gamma}}{1 - \frac{\beta_1}{\gamma\beta_2}} Q_{ij} + \frac{2K\Gamma + \kappa\frac{\beta_1}{\beta_2}}{\gamma \kappa\left(1 - \frac{\beta_1}{\gamma\beta_2}\right)} \frac{\mathrm{D}}{\mathrm{D}t}Q_{ij} \quad .
\end{align}
We find that the choice $\tau_d = 1/[\gamma\kappa(1 - \beta_1/(\gamma \beta_2))]$ and $\tau = [1 - \beta_1/(\gamma\beta_2)]/[2K(\Gamma - 1/(\beta_2^2\gamma))]$ allows to identify Eq. \ref{eq:passiveTissue} with Eq. \ref{eq:diffR}. The last term on the right-hand side of Eq. \ref{eq:passiveTissue},  involving the derivative of $Q_{ij}$, is not present in Eq. \ref{eq:diffR}. The dimensionless prefactor in front of this term, $\alpha=[2K\Gamma + \kappa\beta_1/\beta_2]/[\gamma\kappa(1 - \beta_1/(\gamma\beta_2))]$, can be set to $0$ if $\beta_1= -2K\Gamma \beta_2/\kappa$. This choice implies the following inequality
\begin{align}
\begin{split}
\frac{\tau}{\tau_d}&= \frac{\left[2K\Gamma +\gamma\kappa\right]^2}{2K\left(\Gamma -\frac{1}{\gamma\beta_2^2}\right) \gamma\kappa}\geq \frac{\left[2K\left(\Gamma - \frac{1}{\gamma\beta_2^2}\right) +\gamma\kappa \right]^2}{2K\left(\Gamma -\frac{1}{\gamma\beta_2^2}\right) \gamma\kappa}\geq 4 \quad, 
\end{split}
\end{align}
where we have used $\gamma, \kappa, K, \Gamma \geq 0$ and $\Gamma - 1/(\gamma \beta_2^2) \geq 0$. Therefore, if the parameter $\beta_1$ is chosen such that $\alpha$ is negligible, the gel can never be in the oscillatory regime $\tau_d > \tau/4$ described in Section \ref{sec:tissueShearRheology}.

A finite value of $\beta_1$ implies a dependence of the shear stress on the molecular field $H_{ij}$. If we assume that the main contribution to the shear stress comes from the viscous and elastic stresses and other contributions are negligible, one would set $\beta_1 = 0$. In this case it is no longer possible for $\alpha$ to be arbirtrarily small. Indeed, due to the positive semi-definiteness of Onsager coefficient matrix $\Gamma - 1/(\gamma \beta_2) \geq 0$, the coefficient $\Gamma$ cannot be arbitrarily small. For $\Gamma < 1/(\gamma \beta_2^2)$ the timescale $\tau$ would become negative and the thermodynamic equilibrium state would become unstable. Moreover, a lower limit to the prefactor $\alpha$ is given by 
\begin{align}
\alpha > \frac{\tau_d}{\tau} \quad .
\end{align}
Other possible choices of parameters which would allow neglecting the term involving the shear strain are $K \rightarrow 0$ and $\gamma \kappa \rightarrow \infty$. However, both choices inevitably remove a term in Eq. (\ref{eq:passiveTissue}), in such a way that the form of Eq. (\ref{eq:diffR}) can not be reproduced. Therefore, a passive gel with only viscous and elastic stresses can reproduce Eq. (\ref{eq:diffR}) only with an additional term involving time derivative of the tensor $Q_{ij}$.

\subsection{Active gel}
If the system is provided with a reservoir of ATP molecules to keep it out of equilibrium, such that the difference of chemical potential $\Delta \mu $ is not 0, the equation for $R_{ij}$ in the case $\beta_1 = 0$ reads
\begin{align}
\left(1 + \frac{1}{\gamma \kappa - \theta\Delta\mu} \frac{\mathrm{D}}{\mathrm{D}t}\right) R_{ij} &= \left(2K\Gamma -\overline{\psi}\Delta\mu -\left(\frac{2K}{\beta_2} - \overline{\theta}\Delta\mu\right)\frac{\kappa/\beta_2 - \psi \Delta\mu}{\gamma\kappa - \theta\Delta\mu}\right) Q_{ij} \nonumber\\
&+  \frac{2 K \Gamma -\overline{\psi}\Delta\mu}{\gamma \kappa - \theta\Delta \mu} \frac{\mathrm{D}}{\mathrm{D}t}Q_{ij} 
\end{align}
and we identify 
\begin{align}
\tau_d&= \frac{1}{\gamma\kappa - \theta\Delta \mu}\\
\tau&= \frac{1}{\left(2K\Gamma -\overline{\psi}\Delta \mu- \left(\frac{2K}{\beta_2} - \overline{\theta}\Delta\mu\right)\frac{\kappa/\beta_2 - \psi\Delta\mu}{\gamma\kappa - \theta \Delta\mu}\right)} \quad .
\end{align}
In this case, the term involving the derivative of shear strain $Q_{ij}$ becomes small when $\left|(2K\Gamma - \overline{\psi}\Delta\mu)/(\gamma \kappa - \theta \Delta\mu)\right| \ll 1$. The stability in this limit can still be maintained if the factor $(2K/\beta_2 -\overline{\theta}\Delta\mu)(\kappa/\beta_2 - \psi\Delta\mu)$ is negative. This is possible because the signs of the active terms $\psi\Delta\mu$ and $\overline{\theta}\Delta\mu$ are not constrained. Thus, an active nematic gel close to equilibrium can be described by Eq. \ref{eq:diffR}, even without a specific coupling of molecular field to the shear stress.

\section{Autonomous convergent extension}\label{appendixConvergentExtension}
Here we solve in detail the example from the \sref{sec:activityFlows}. We consider a rectangular homogeneous tissue with length $l$ and height $h$. The cell polarity nematic $q_{ij}$ is constant and oriented along x-axis such that $q_{xx} = 1, q_{xy} = 0$. We assume that for $t<0$, the tissue is at rest, not subjected to active anisotropic processes, that cells are isotropic and that there is no stress in the elastic material surrounding the tissue. At $t=0$, active anisotropic processes are turned on and remain constant for $t>0$. The tissue is constrained in space by surrounding elastic material which provides resistance to the changes in tissue length and height. We describe the external elastic response to the tissue deformations by a Hooke's law
\begin{align}\label{eq:boundaryConditions}
\begin{split}
\sigma_{xx} &= -k_x L\\
\sigma_{yy} &= -k_y H \quad, 
\end{split}
\end{align}
where $L = \ln{(l/l_0)}$ and $H = \ln{(h/h_0)}$ are natural strain variables of the tissue and $l_0$, $h_0$ are tissue length and width at zero stress. For negative times, the tissue is stress free and thus $L(t < 0)= H(t< 0) = 0$.

Since the shape of the tissue is constrained to be a rectangle, the velocity gradient tensor has only diagonal components:
\begin{align}\label{eq:rectangleVelGradComponents}
\begin{split}
v_{xx} &= \partial_t L\\
v_{yy} &= \partial_t H \quad .
\end{split}
\end{align}

Combining these relations with the constitutive relation for the isotropic stress in Eq. \ref{eq:isotropicStressConstitutive}, we obtain a relation between $L$ and $H$ from the isotropic flow component
\begin{align}\label{eq:rectIso}
\left(1 + \frac{k_x}{2\overline{K}}\right) \partial_t L + \left(1 + \frac{k_y}{2\overline{K}}\right) \partial_t H &= 0 \quad ,
\end{align}
where we have used the fact that in the absence of cell division and extrusion, $(da/dt)/a=(dA/dt)/A$ with $A$ being the area of the tissue.
Using the constitutive relation for the shear stress Eq. \ref{eq:fullStressConstitutive}, the constitutive equation for the shear due to topological rearrangements Eq. \ref{eq:fullR}, and the shear decomposition Eq. \ref{eq:shearDecomposition}, we obtain a second relation between $L$ and $H$:
\begin{align}\label{eq:rectAniso}
\begin{split}
\left(1 + \frac{k_x}{2K}\right)\left(1 + \tau_d \partial_t\right)\partial_t L + \frac{k_x}{2K\tau}L 
-\left(1 + \frac{k_y}{2K}\right)\left(1 + \tau_d \partial_t\right)\partial_t H - \frac{k_y}{2K\tau}H= 2\gamma
\end{split}
\end{align}
where $\gamma= -\zeta/(2\eta) + \lambda$.
Integrating Eq. \ref{eq:rectIso} from an arbitrary lower bound $t< 0$  yields
\begin{align}
\label{LHrelation}
  \left(1 + \frac{k_x}{2\overline{K}}\right) L + \left(1 + \frac{k_y}{2\overline{K}}\right) H &= 0 \quad .
\end{align}
We now express $H$ in terms of $L$ in Eq. \ref{eq:rectAniso} and we obtain the second order equation
\begin{align}\label{eq:singleLFull}
\tau_d \partial^2_tL + \partial_tL + \frac{\nu}{\tau\mu} L &= \frac{2\gamma}{\mu} \quad ,
\end{align}
where
\begin{align}
\label{Definitionmu}
\mu &= 1 + \frac{k_x}{2K} + \left(1 + \frac{k_y}{2K}\right)\frac{1 + \frac{k_x}{2\overline{K}}}{1 + \frac{k_y}{2\overline{K}}}\\
\label{Definitionnu}
\nu &= \frac{k_x}{2K} + \frac{k_y}{2K}\frac{1 + \frac{k_x}{2\overline{K}}}{1 + \frac{k_y}{2\overline{K}}} \quad .
\end{align}
This can be solved for
\begin{align}\label{eq:solutionL}
\begin{split}
L(t) &= \frac{2\tau}{\nu}\gamma\left[1 - \left(1 - \frac{\nu L(0)}{2\tau\gamma}\right)e^{-t/(2\tau_d)}\left(\frac{1}{s}\sin{\frac{s}{2\tau_d}t} + \cos{\frac{s}{2\tau_d}t}\right)\right] \\
&+ \partial_tL(0) e^{-t/(2\tau_d)}\frac{\sin{\frac{s}{2\tau_d}t}}{s/(2\tau_d)} \quad ,
\end{split}
\end{align}
where $L(0)$ and $\partial_tL(0)$ are initial conditions, and 
\begin{align}\label{eq:parameter_sr}
s^2 &= \frac{4\tau_d\nu}{\tau \mu} - 1.
\end{align}
The parameter $s^2$ is positive only for high enough values of the memory timescale $\tau_d$. When $s^2$ becomes negative, the solution is equivalent in form to Eq. \ref{eq:solutionL}, with trigonometric functions replaced by their hyperbolic counterparts and $s^2$ replaced with $-s^2$.

 We now determine the initial conditions for an experimental setting in which the tissue is initially at rest, cells in the tissue are not elongated, external elastic connections are not under tension, and there is no tissue polarity. At $t= 0$ the polarity is activated on a timescale much shorter than $\tau$ and $\tau_d$, so that we can treat the activation as instantaneous. First, considering Eq. \ref{eq:fullR} we can conclude that $R_{xx}(0^+) = 0$. Then, using Eqs. \ref{eq:boundaryConditions}, \ref{eq:rectangleVelGradComponents}, \ref{eq:rectIso}, \ref{eq:shearDecomposition} and \ref{eq:fullStressConstitutive} we can show that for $t<0^+$
\begin{align}
\begin{split}
\frac{1}{2}\left(1 + \frac{1 + \frac{k_x}{2\overline{K}}}{1 + \frac{k_y}{2\overline{K}}}\right)\partial_tL(t) &= \partial_t Q_{xx}(t)\\
&= -\frac{1}{2}\left[\frac{k_x}{2K} + \frac{k_y}{2K}\frac{1 + \frac{k_x}{2\overline{K}}}{1 + \frac{k_y}{2\overline{K}}}\right]\partial_tL(t) - \frac{\zeta}{2K}\partial_t q_{xx} (t)\quad .
\end{split}
\end{align}
Then, using $\partial_t q_{xx} = \delta(t)$ and integrating over a small finite time interval around $t= 0$, we obtain the initial conditions
\begin{align}
\begin{split}\label{eq:initCondL}
L(0^+) &= -\frac{\zeta}{\mu K}\\
\partial_tL(0^+)&= 0 \quad .
\end{split}
\end{align}

When the elastic connections around the tissue are isotropic ($k_x=k_y$), one obtains from Eqs. \ref{LHrelation}, \ref{Definitionmu} and \ref{Definitionnu}:
\begin{align}\label{eq:isotropicBoundarySprings}
H(t) &= - L(t)\\
\mu &= 2\left(1 + \frac{k}{2K}\right)\\
\nu &= \frac{k}{K} \quad .
\end{align}
Since $L(t) + H(t) = 0$, the tissue area is constant. Therefore, area changes arise only when the surrounding material is anisotropic. Using these relations, the initial conditions discussed above and Eq. \ref{eq:solutionL}, we obtain Eq. \ref{eq:exampleSolutionL} in the main text.

Finally, we note that in the limit $\tau_d \to 0$, Eq. \ref{eq:solutionL} describes a simple exponential relaxation of the tissue length:
\begin{align} \label{eq:limitTauD}
L(t) &=\frac{2\tau\gamma}{\nu}\left[1 - \left(1 - \frac{\nu L(0)}{2\tau\gamma}\right) e^{-\frac{\nu}{\mu \tau}t}\right] \quad .
\end{align}
with a characteristic relaxation timescale equal to $(k \tau)/(2K + k)$ for isotropic external springs. It differs from the Maxwell timescale of the tissue $\tau$ by a factor describing a competition of internal and external elasticities.

\section{Couette flow in active tissue}\label{app:Couette}
Here we consider a two-dimensional tissue, which is constrained on two sides by straight boundaries, setting the tissue width to a fixed value $h$. The tissue is fixed on its lower side and it is translationally invariant along the boundaries:
\begin{align}
v_x(0)&= 0\\
v_y(0)&= 0\\
v_y(h)&= 0\\
\partial_x F &= 0
\end{align}
for any quantity $F$. For simplicity we assume that the tissue is incompressible
\begin{align}
\partial_x v_x+\partial_yv_y&= 0 \quad ,
\end{align}
so we can conclude that $v_y= 0$. Force balance in this system reads in the absence of external force
\begin{align}\label{eq:x_forceBalance}
\partial_y \sigma_{xy}&= 0,\\
\partial_y \sigma_{yy}&= 0.
\end{align}
From Eqs. \ref{eq:shearDecomposition} and \ref{eq:fullR} we find in the frequency domain
\begin{align}
\label{eq:xy_shear_decomposition} \frac{1}{2}\partial_y v_x&= \frac{1}{\tau}\left[i \omega \tau + \frac{1}{1 + i\omega \tau_d}\right] Q_{xy} + \frac{1}{1 + i\omega\tau_d}\lambda q_{xy} \quad .
\end{align}
where for simplicity we have neglected non-linear terms coming from corotational derivatives. Combining Eqs. \ref{eq:stressConstitutive}, \ref{eq:x_forceBalance} and \ref{eq:xy_shear_decomposition} we obtain the equation for $v_x$
\begin{align}\label{eq:v_x_dynamics}
\partial_{y}^2v_{x} &= -2\left[i \omega \tau + \frac{1}{1 + i\omega \tau_d}\right]\frac{\partial_y\left(\zeta q_{xy}\right)}{2K\tau}  + \frac{2}{1+ i\omega\tau_d}\partial_y\left(\lambda q_{xy}\right).
\end{align}
We consider the case when active terms $\lambda q_{xy}$ and $\zeta q_{xy}$  are homogeneous is space so that the right hand side vanishes of Eq. \ref{eq:v_x_dynamics} vanishes. The solution of Eq. \ref{eq:v_x_dynamics} is then
\begin{align}
v_x(y, \omega)&= v_x(h, \omega)\frac{y}{h} \quad ,
\end{align}
and the stress $\sigma_{xy}$, which is constant in space, can be evaluated at $y= h$ to be
\begin{align}
\sigma_{xy}(h, \omega)&= \chi_h(\omega)\left[v_x(h, \omega) - V_q(\omega)\right] \quad .
\end{align}
Here, the response function
\begin{align}
\chi_h(\omega)&= \frac{K\tau}{h} \frac{1 + i\omega \tau_d}{i\omega\tau \left(1 + i\omega\tau_d\right) + 1}
\end{align}
is proportional to the response function in Eq. \ref{eq:freqRheology}, and the spontaneous velocity is given by
\begin{align}
V_q(\omega)&= 2h\left[-i \omega \frac{\zeta}{2K}q_{xy} + \frac{1}{1 + i\omega\tau_d}\left(\lambda - \frac{\zeta}{2K\tau}\right)q_{xy}\right]
\end{align}
 If the top tissue boundary at $y= h$ is free to slide, $\sigma_{xy}(h)= 0$ and $v(h,\omega)=V_q(\omega)$. In that case we find the following expression of the tissue velocity as a response to the cell polarity: 
\begin{align}
v_x(h, t)&= -\frac{h}{K}\partial_t\left(\zeta q_{xy}\right) + 2 h \int\limits_{-\infty}^t\frac{1}{\tau_d}e^{-\frac{1}{\tau_d}\left(t - t'\right)}  \left(\lambda - \frac{\zeta}{2K\tau}\right)q_{xy}dt' \quad .
\end{align}
For a immobile tissue starting with zero values of $\lambda q_{xy}$ and $\zeta q_{xy}$ for $t<0$ and constant values for $t>0$ obtain Eq. \ref{eq:CouetteMain}.

\vspace{1.2in}

\bibliographystyle{unsrt}
\bibliography{main.bib}

\end{document}